\documentclass[twocolumn]{aastex631}

\newcommand{\A}{\AA~}

\def\nus{{\it NuSTAR }}
\def\nic{{\it Nicer }}
\def\nus{{\it NuSTAR }}
\def\sw{{\it Swift}}
\def\fer{{\it Fermi}}

\shorttitle{Spin-down accretion regime of ULXP Swift J0243 }
\shortauthors{Liu et al.}

\begin{document}
\title{The spin-down accretion regime of Galactic ultra-luminous X-ray pulsar Swift J0243.6+6124}
\correspondingauthor{Jiren Liu}
\email{liujiren@bjp.org.cn}
\author{Jiren Liu}
\affiliation{Beijing Planetarium, Beijing Academy of Science and Technology, Beijing 100044, China}

\author{Long Ji}
\affiliation{School of Physics and Astronomy, Sun Yat-sen University, 2 Daxue Road, Zhuhai, Guangdong 519082, China}
\author{Mingyu Ge}
\affiliation{Institute of High Energy Physics, Chinese Academy of Sciences,
    Beijing 100049, China}

%


\begin{abstract}
The relative high fluxes of the Galactic ultra-luminous X-ray pulsar 
Swift J0243 allow a detailed study of its spin-down regime in quiescence state,
for a first time.
After the 2017 giant outburst, its spin frequencies show a 
linear decreasing trend with some variations due to minor outbursts.
The linear spin-down rate is $\sim-1.9\times10^{-12}$ Hz/s during the 
period of lowest luminosity, from which one can 
infer a dipole field $\sim1.75\times10^{13}$ G.
The $\dot{\nu}-L$ relation during the spin-down regime is complex, and the 
$\dot{\nu}$ is close to 0 when the luminosity reaches 
both the high end ($L_{38}\sim0.3$) and the lowest value ($L_{38}\sim0.03$).
The luminosity of zero-torque is different for the giant outburst 
and other minor outbursts. It is likely due to different 
accretion flows for different types of outburst, as evidenced by the differences of the
spectra and pulse profiles at a similar luminosity for different types of 
outburst (giant or not).
The pulse profile changes from double peaks in the spin-up state
to a single broad peak in the low spin-down regime, indicating 
the emission beam/region is larger in the low spin-down regime.
These results show that accretion is still ongoing in the low spin-down regime for 
which the neutron star is supposed to be in a propeller state.

\end{abstract}


\section{Introduction}

Accretion-powered X-ray pulsars are strongly magnetized neutron stars 
that accrete material from normal companion stars. 
The accreting material also contains angular momentum that changing the 
spin period of the neutron star.
The spin-up/spin-down behavior of the neutron star depends on the interaction 
between the accreting flow and the magnetosphere of the neutron 
star \citep[e.g.][]{RJ77,GL79, Wang95}.
Therefore, the observed spin behavior can be used to constrain 
the interaction of the accreting flow with the magnetosphere of X-ray pulsars.

Be-type X-ray binaries (BeXBs) are ideal targets for the study of the spin behavior 
of X-ray pulsars since they cover a large dynamical range of luminosity (accretion rate).
In BeXBs, the Be/Oe-type donor star rotates rapidly and forms an equatorial 
decretion disk,
and the neutron star, generally in an eccentric orbit, will produce a normal outburst
(type I, with peak luminosities below $10^{37}$erg\,s$^{-1}$)
around the periastron passage. Occasionally, they can produce a giant outburst (type II),
which could occur at any orbital phase and reach a peak luminosity higher than 
$10^{38}$erg\,s$^{-1}$ \citep[e.g.][]{Rei11}. 
During the outbursts, the pulsars generally undergo spin-up episodes, the rates of which
correlate with luminosity \citep[e.g.][]{Bil97}. In-between the outbursts, the pulsars 
are generally in a quiescence state and a spin-down trend is observed.

Due to the relatively low luminosity, the spin-down accretion regime of BeXBs is 
difficult to monitor and is rarely explored
compared with the spin-up regime. The spin-down accretion regime, however, is critical
to reveal the interaction of the accreting flow with the magnetosphere around 
the co-rotation radius, beyond which the threading of the disk by the magnetic field 
applies a negative torque to the neutron star. In this paper, we study the 
spin-down regime of the Galactic ultra-luminous X-ray pulsar Swift J0243.6+6124
(hereafter J0243), the quiescence state of which is bright enough to provide 
a unique opportunity to study the spin-down accretion regime, for a first time.

\begin{figure*}
	\includegraphics[width=6.0in]{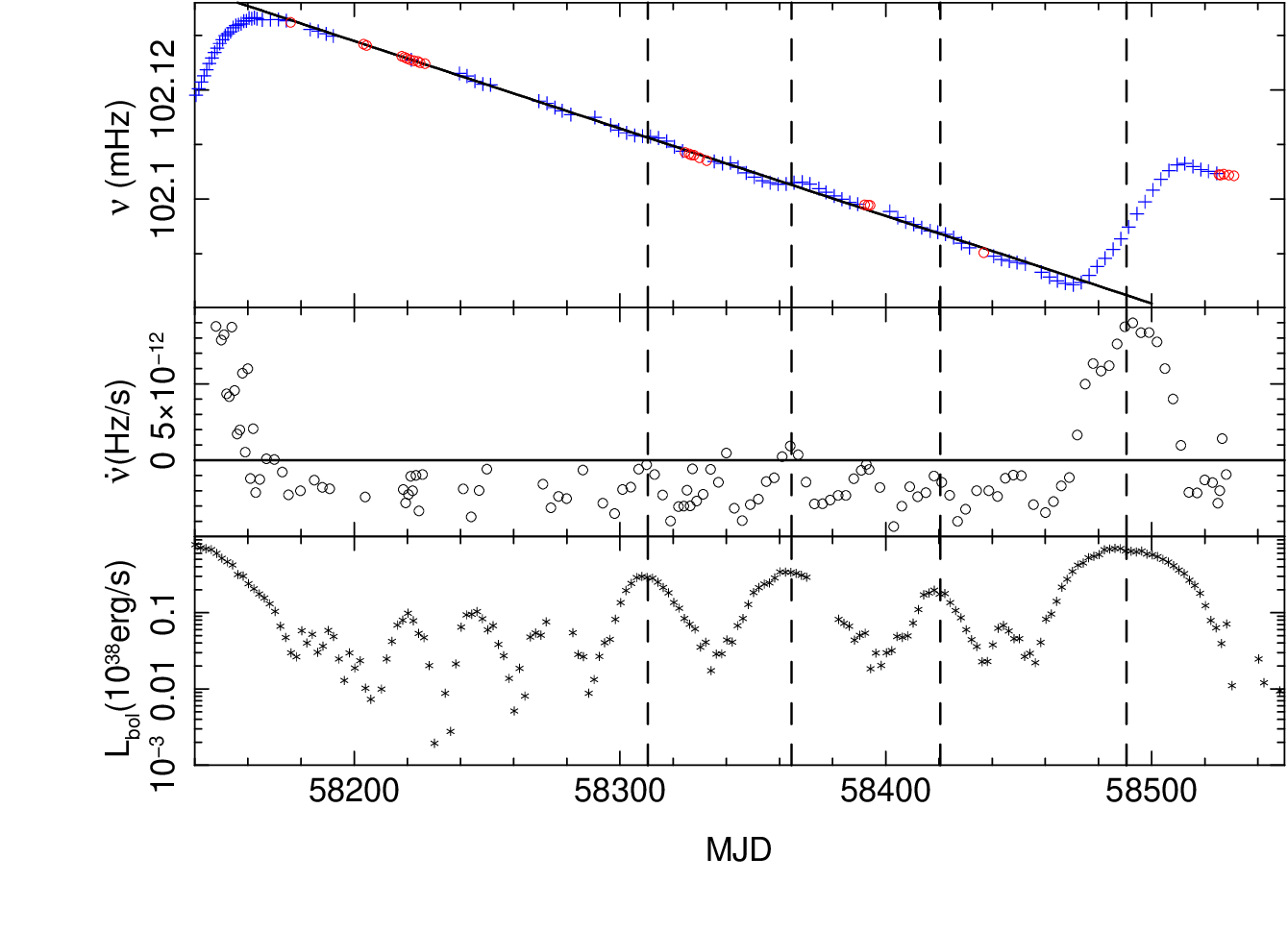}
	\caption{The spin frequency, spin derivative, and bolometric luminosity of Swift J0243 
during the giant outburst in 2017. A 300-days long spin-down period
is present after the giant outburst, with a last outburst near MJD 58500.
The red circles in the top panel are measurements from {\it Nicer} data.
The four vertical dashed lines (around MJD 58310, 58365, 58420, 58490)
indicate the four later outbursts.
} 
\end{figure*}

Swift J0243 is a new BeXB discovered by Swift in October 2017 
during one of the brightest outburst \citep{Cen17}.
At a distance of 5.2 kpc \citep{Bai21}, 
its peak luminosity reaches $\sim10^{39}$erg\,s$^{-1}$, which makes it 
the first Galactic ultra-luminous X-ray pulsar \citep[e.g.][]{Tsy18,Wil18}.
We note that a larger distance of 6.8 kpc were generally adopted in 
earlier studies as measured from {\it Gaia} data release 2 \citep{Bai18}.
After the giant outburst, it underwent several minor outbursts and one 
last relatively bright outburst around Jan. 2019.
Between the giant outburst and the last brighting, a long spin-down trend was 
observed (Figure 1).

\section{Observation data}

The 2017 giant outburst of Swift J0243 has been monitored by many
existing X-ray instruments, including {\it Fermi}, {\it Swift}, {\it Nicer},
{\it MAXI}, and {\it Insight-HXMT}. 
The Gamma-ray Burst Monitor \citep[GBM,][]{GBM09} on-board the {\it Fermi} spacecraft 
provides a continuous monitoring of the spin history of X-ray
pulsars\footnote{https://gammaray.msfc.nasa.gov/gbm/science/pulsars/}
\citep[e.g.][]{Fin09,Mal20}. The spin evolution of J0243 measured by GBM is presented
in Figure 1, together with the bolometric luminosity converted from 
the \sw/BAT flux in 15-50 keV (see next section). 
For completeness, we added a few measurements from \nic data
(red circles in Figure 1) when the spin 
frequency was not detected by \fer/GBM.
The spin derivatives, calculated from two contiguous measurements, 
are plotted in the middle panel.
As can be seen, most of the spin derivatives were negative after the 
giant outburst (later than MJD 58160), except for the 
last brighting around MJD 58500.

\begin{figure}
	\includegraphics[width=3.3in]{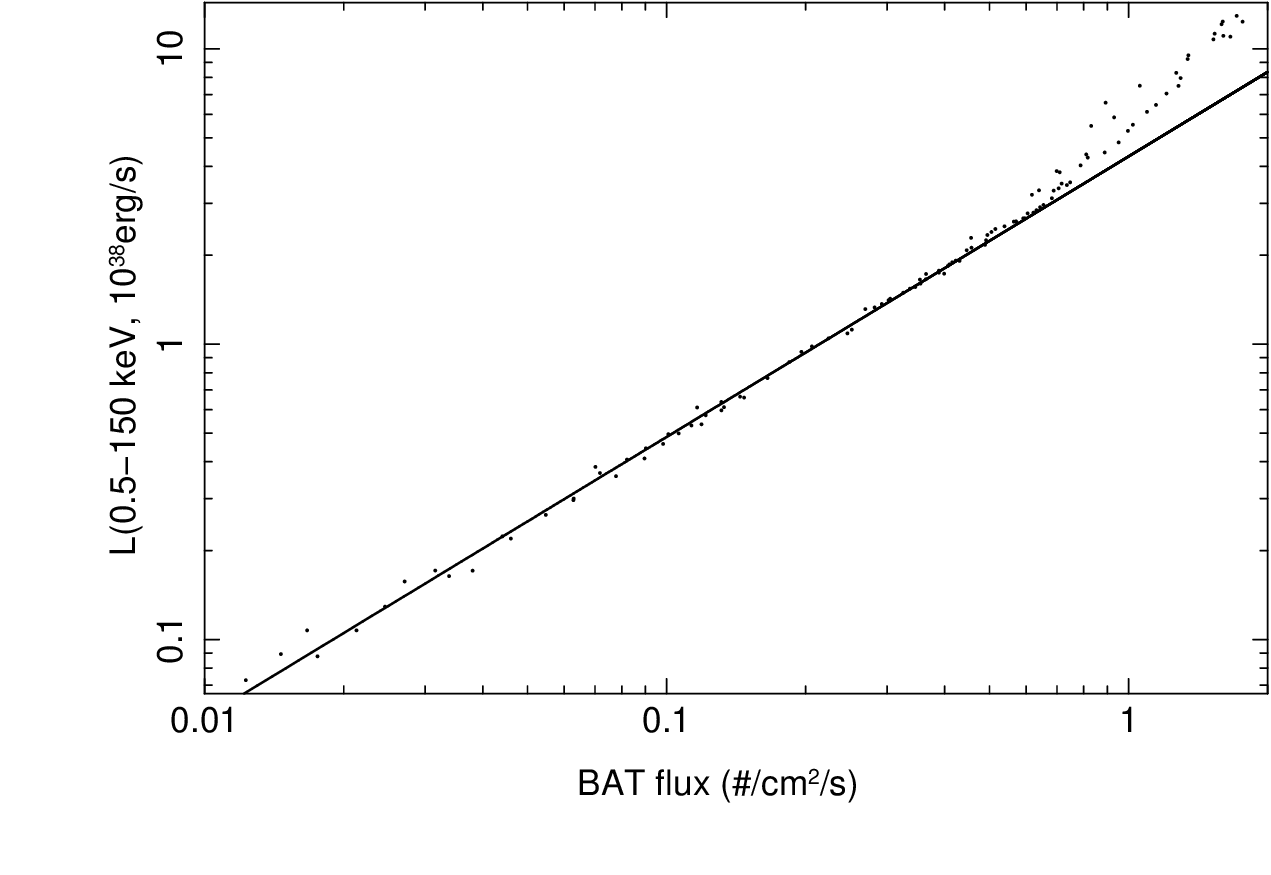}
	\caption{Comparison of the HXMT-estimated bolometric luminosities 
of Swift J0243 with the BAT fluxes. 
} 
\end{figure}

To convert the BAT flux to bolometric luminosity, we compare the 0.5-150 keV 
luminosities measured by HXMT with the BAT fluxes at the corresponding HXMT 
observation time.
The luminosities of HXMT observation were taken from \citet{Liu22}.
As shown in Figure 2, below $2\times10^{38}$erg\,s$^{-1}$, 
the BAT fluxes and HXMT luminosities follow a relation of
$L_{38}=4.3\times f_{BAT}^{0.95}$, where $L_{38}$ is the bolometric luminosity
in units of $10^{38}$erg\,s$^{-1}$. 
So we convert the BAT flux to bolometric luminosity with this relation.

\section{Results}

\subsection{Spin-down behavior}

As can be seen from Figure 1, after the giant outburst, 
the spin frequencies show a linear decreasing trend, 
with some variations due to the middle minor outbursts.
We fit a linear function to the data points within MJD 58190 and 58290,
where the fluxes are relatively low (the averaged luminosity 
is about $3\times10^{36}$ erg\,s$^{-1}$). 
We obtain a spin-down rate of $-1.9\times10^{-12}$ Hz/s, which
is plotted as the solid line in the top panel of Figure 1.

Besides the linear decreasing trend, we observe two kinds of fluctuations
of the spin-down rates. One is around the peak of the minor outbursts, 
where the spin-down rates increased, or even reversed to spin-up, 
due to the increasing accretion rate, 
as indicated by the vertical dashed lines in Figure 1.
We see another kind of fluctuation from the middle panel of Figure 1. 
Between the flux outbursts, around the flux minima 
(such as around MJD 58335 and 58390),
the spin derivatives are also reaching 0, similar to those 
around outburst peaks. Such a behavior is unexpected.

\begin{figure}
	\hspace{-0.5cm}
	\includegraphics[width=3.5in]{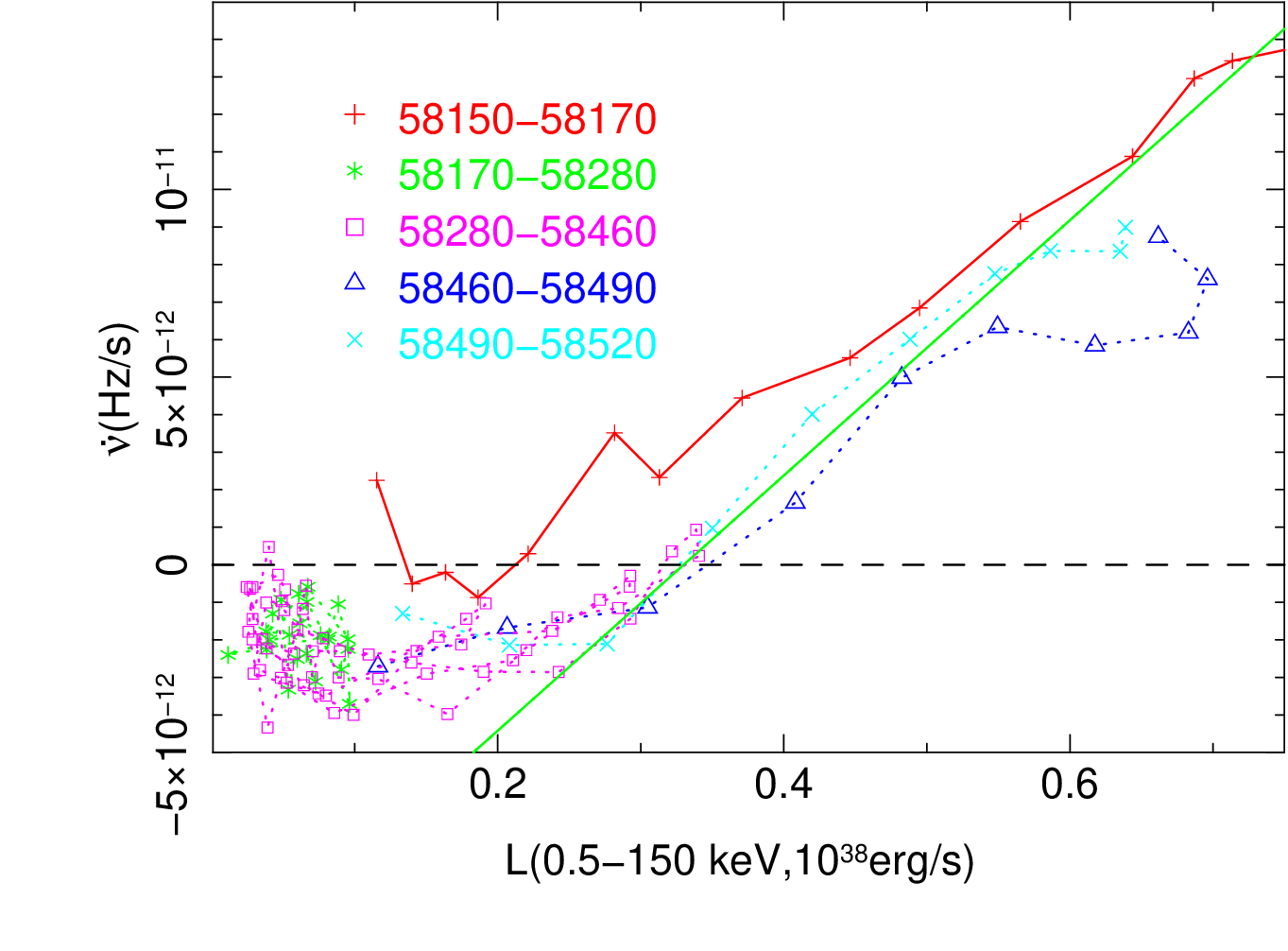}
	\caption{The $\dot{\nu}-L$ relation of Swift J0243 
	observed during different spin-up/spin-down periods.
	A solid green line is drawn to help to guide eyes.
} 
\end{figure}

\subsection{Spin-down derivative vs luminosity}

To quantify the relation between the spin-down derivative and accretion luminosity,
we plot the $\dot{\nu}-L$ relation during different spin-down periods in Figure 3.
The spin derivative $\dot{\nu}$ is calculated from two contiguous measurements,
the time interval between which is smaller than 10 days. The luminosity is calculated
as an averaged value over the period. 
To check the differences of different outbursts and luminosities, we divided the time 
range into MJD 58150-58170 (the fading period of the giant outburst), 
58170-58280 (the relatively low luminosity period), 58280-58460 (the middle 
minor outbursts), 58460-58490 (the growing period of the last outburst) and 
58490-58520 (the fading period of the last outburst).
For the period between MJD 58150 and 58170, 
the calculated $\dot{\nu}$ from GBM data show larger fluctuations and 
we used the $\dot{\nu}$ calculated from HXMT data. 

First, we see that the $\dot{\nu}-L$ relation of the giant outburst 
shows an offset from that of the last outburst around 
MJD 58500. The zero point of $\dot{\nu}=0$ is around 
a luminosity $\sim0.2\times10^{38}$erg\,s$^{-1}$ for the giant outburst, 
while it is around $0.33\times10^{38}$erg\,s$^{-1}$ for the last outburst
and for the minor outburst around MJD 58365.

During the relatively low period between MJD 58170 and 58280, 
the measured $\dot{\nu}$ are scattered around $-2\times10^{-12}$ Hz/s, 
with the luminosities below $0.1\times10^{38}$erg\,s$^{-1}$.
For the period between MJD 58280 and 58460, above 
$0.2\times10^{38}$erg\,s$^{-1}$, the spin-down derivative decreases when the 
luminosity decreases; while below $0.1\times10^{38}$erg\,s$^{-1}$,
the averaged trend of spin-down rate seems to increase when 
the luminosity decreases to the lowest value, as already noted 
in previous section. And there seems to be two zero points 
of $\dot{\nu}=0$: one for $0.33\times10^{38}$erg\,s$^{-1}$ and 
one for $0.03-0.04\times10^{38}$erg\,s$^{-1}$.

For the last outburst between MJD 58460 and 58520, 
for $\dot{\nu}>0$, the slope of the $\dot{\nu}-L$ relation is similar to 
that of the giant outburst; while for $\dot{\nu}<0$, the slope 
becomes flattened when the luminosity is below 
$0.25\times10^{38}$erg\,s$^{-1}$.

\begin{figure}
	\includegraphics[width=3.4in]{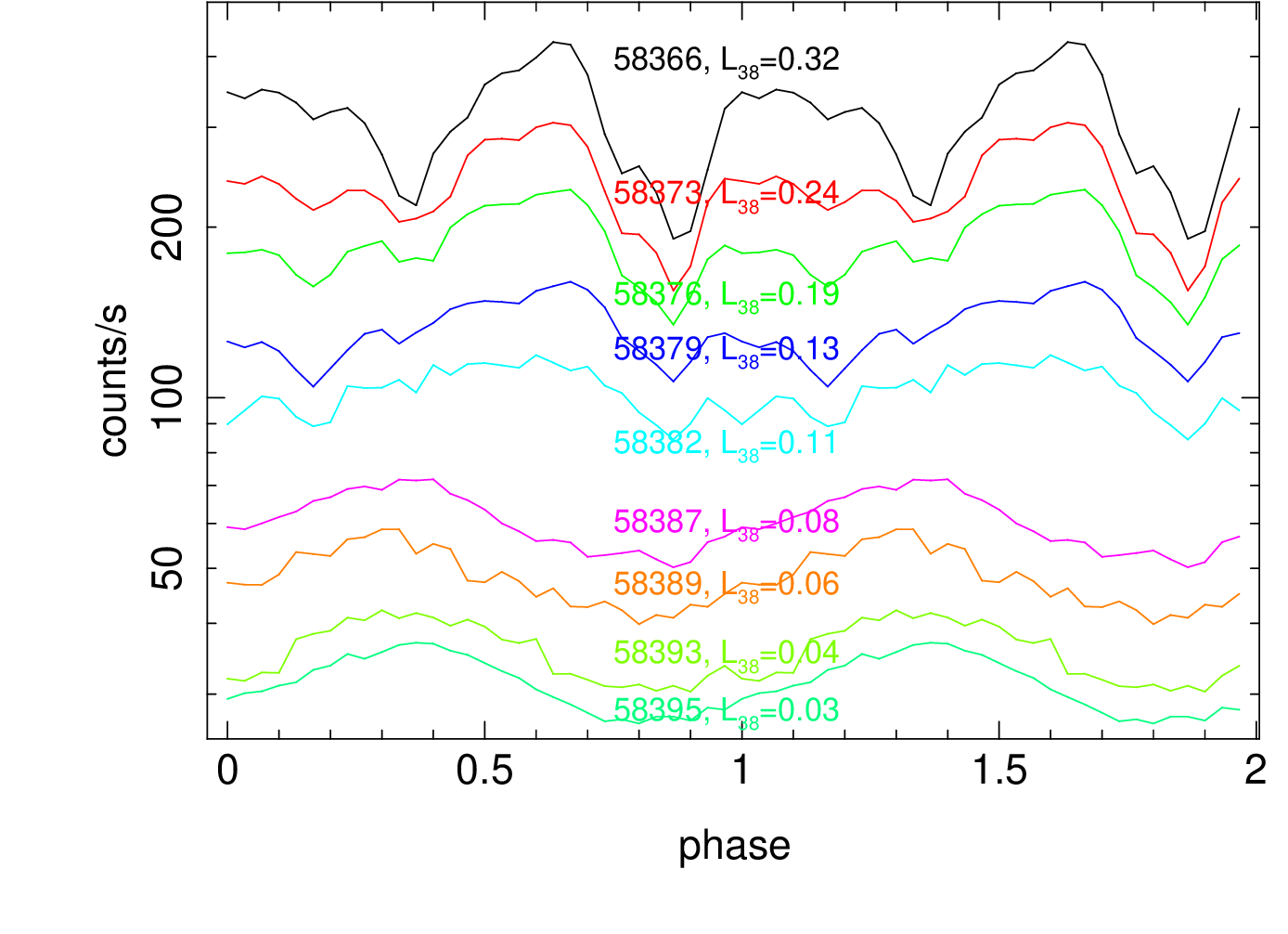}
	\includegraphics[width=3.4in]{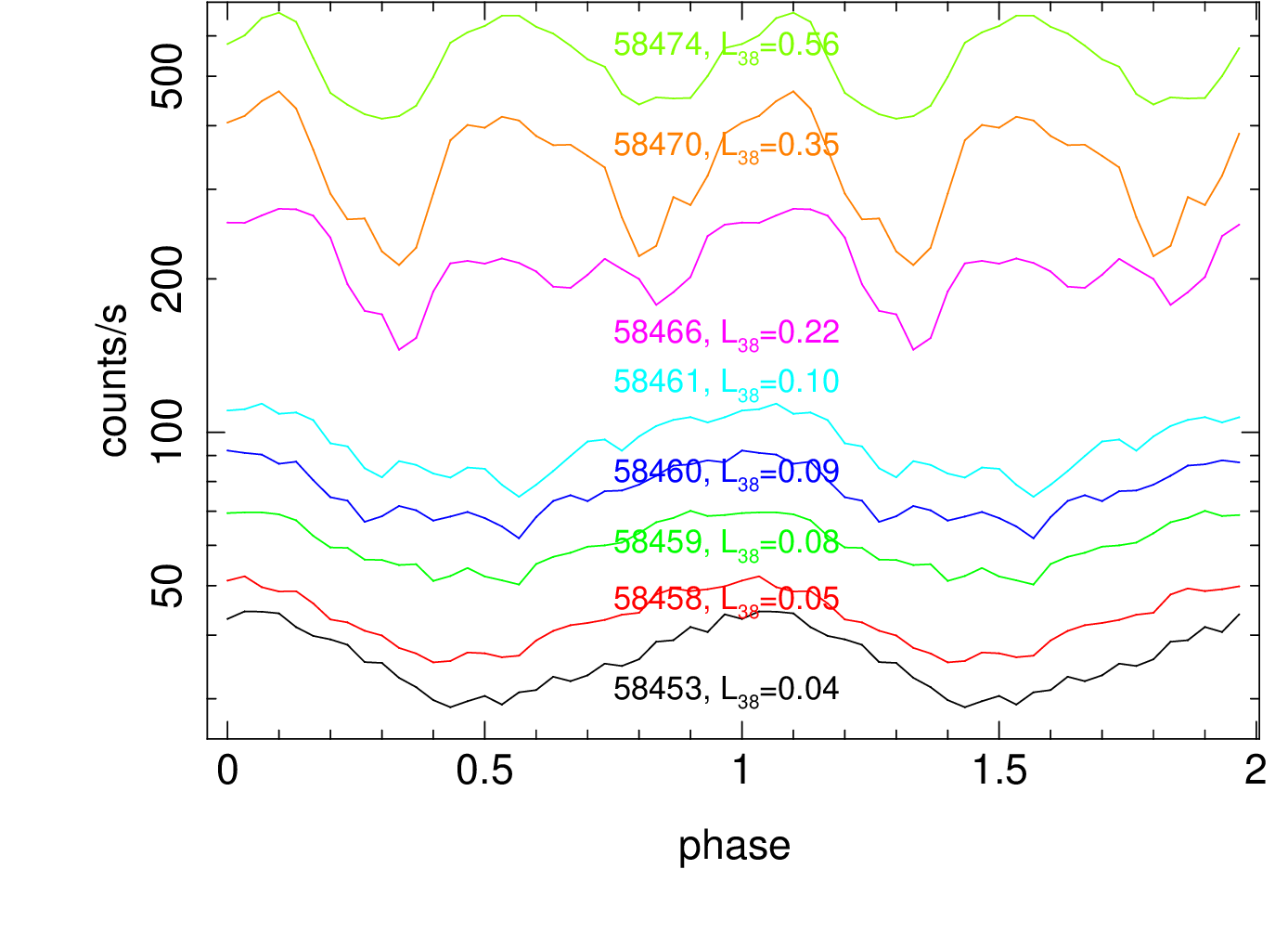}
	\caption{Example of the pulse profile evolution of Swift J0243 during the 
	spin-down/spin-up transition regime within 0.4-8 keV observed by {\it Nicer}.
	The chosen periods are around MJD 58366-58395 (top) and MJD 58453-58474 (bottom).
} 
\end{figure}

\subsection{Evolution of the pulse profile}

The pulse profiles of J0243 had been extensively studied in previous work, but 
no attention was paid for the transition period between spin-up and spin-down 
regimes. 
We find that the profile mostly depends on the luminosity, and
here we choose two periods (MJD 58366-58395 and 58453-58470)
to illustrate the evolution 
of the profile during the spin-down regime. The profile evolution 
during other periods are similar.
The aligned pulse profiles within 0.4-8 keV observed by \nic are plotted in Figure 4.

During the lowest state of luminosity ($L_{38}\sim0.03-0.08$),
the profile is a single broad peak, with a phase width $\sim1$.
Around the luminosity of $L_{38}\sim0.08-0.11$, a minor 
peak appeared left to the main peak. 
During the middle state of luminosity ($L_{38}\sim0.2$),
the main peak splits into one minor peak and one main peak, and 
the phase width of the main peak is about 0.5 in this state.
And the profile looks composed of three peaks.
At the luminosity of $L_{38}\sim0.32-0.35$, when the spin frequency 
changing from spin-down to spin-up, the newly appeared two minor peaks 
merged into one bump, and the profile is composed of double peaks.
At higher luminosity of $L_{38}\sim0.56$, the profile is double-peaked, similar 
to that around $L_{38}\sim0.32-0.35$.

That is, the profile of Swift J0243 changes from a single broad peak to double peaks
when it changes from the lowest luminosity state of spin-down to 
the spin-up state, and in-between, the profile shows a transitional shape.
It indicates the change of accretion geometry/emission region 
during the spin-down/spin-up transition.

\begin{table}
\scriptsize
	\begin{center}
\caption{Fitting parameters of cutoff power-law model}
\begin{tabular}{cccccc}
 \hline
    MJD & $L_{38}$ & $N_H^a$ & Norm$^b$ & $\Gamma$ & E$_{c} (\mbox{keV})$\\
   \hline
   58494 &  0.6 & $0.62\pm0.01$ & $0.464\pm0.004$ & $0.80\pm0.01$&  $15.5\pm0.1$ \\
   58500 &  0.5 & $0.59\pm0.01$ & $0.375\pm0.003$ & $0.70\pm0.01$&  $13.8\pm0.1$ \\
   58506 &  0.4 & $0.64\pm0.01$ & $0.293\pm0.003$ & $0.62\pm0.01$&  $12.5\pm0.1$ \\
   58510 &  0.3 & $0.62\pm0.01$ & $0.221\pm0.003$ & $0.56\pm0.01$&  $11.5\pm0.1$ \\
   58516 &  0.2 & $0.78\pm0.04$ & $0.115\pm0.004$ & $0.50\pm0.02$&  $10.5\pm0.3$ \\
   58520 &  0.1 & $0.75\pm0.06$ & $0.078\pm0.004$ & $0.51\pm0.04$&  $9.9\pm0.4$ \\
   58521 &  0.07 & $0.72\pm0.08$ & $0.057\pm0.004$ & $0.51\pm0.06$&  $9.4\pm0.6$ \\
   58523 &  0.05 & $0.83\pm0.12$ & $0.048\pm0.005$ & $0.65\pm0.08$&  $10.8\pm1.0$ \\
   \hline
   58151 &  0.4 & $0.43\pm0.02$ & $0.277\pm0.005$ & $0.59\pm0.01$&  $11.7\pm0.2$ \\
   58506 &  0.4 & $0.64\pm0.01$ & $0.293\pm0.003$ & $0.62\pm0.01$&  $12.5\pm0.1$ \\
   58157 &  0.25 & $0.45\pm0.03$ & $0.145\pm0.005$ & $0.46\pm0.03$&  $10.8\pm0.3$ \\
   58513 &  0.25 & $0.67\pm0.02$ & $0.140\pm0.002$ & $0.41\pm0.02$&  $10.1\pm0.2$ \\
   58165 &  0.13 & $0.51\pm0.04$ & $0.094\pm0.004$ & $0.50\pm0.03$&  $10.2\pm0.4$ \\
   58519 &  0.13 & $0.74\pm0.04$ & $0.089\pm0.003$ & $0.47\pm0.03$&  $9.6\pm0.3$ \\
   \hline
\end{tabular}
\begin{description}
  \begin{footnotesize}
  \item $^a$ $N_H$ is the hydrogen column density in units of $10^{22}$ cm$^{-2}$;
  \item $^b$ Norm is the normalization of the power-law model at 1 keV, in units of
photons keV$^{-1}$\,cm$^{-2}$\,s$^{-1}$.
  \end{footnotesize}
   \end{description}
\end{center}
\end{table}

\subsection{Evolution of the spectrum}

To study the spectral evolution of Swift J0243 during the spin-down regime,
we use the HXMT observation of the last outburst, which covered MJD 58493 
to 58523. These observations are almost on a daily base and have a wide 
energy coverage. 
HXMT has three collimated instruments sensitive to different energy bands: 
LE (1-10 keV), ME (10-30 keV), and HE (25-250 keV), with effective 
areas of 384, 952, and 5100 cm$^2$, respectively \citep{Zhang20}.
The spectra are extracted using the HXMT Data Analysis software v2.04, 
with the calibration model of v2.05 and only use the data from LE and ME.
The spectra are binned with a minimum signal-to-noise of 10.
Some examples of the spectra at different times and 
luminosities are presented in Figure 5. All the spectra can be 
well fitted with an absorbed cutoff-power-law model, with a reduced $\chi^2$ 
smaller than 1.2. The fitting results are listed in Table 1.

\begin{figure}
	\hspace{-0.5cm}
	\includegraphics[width=3.5in]{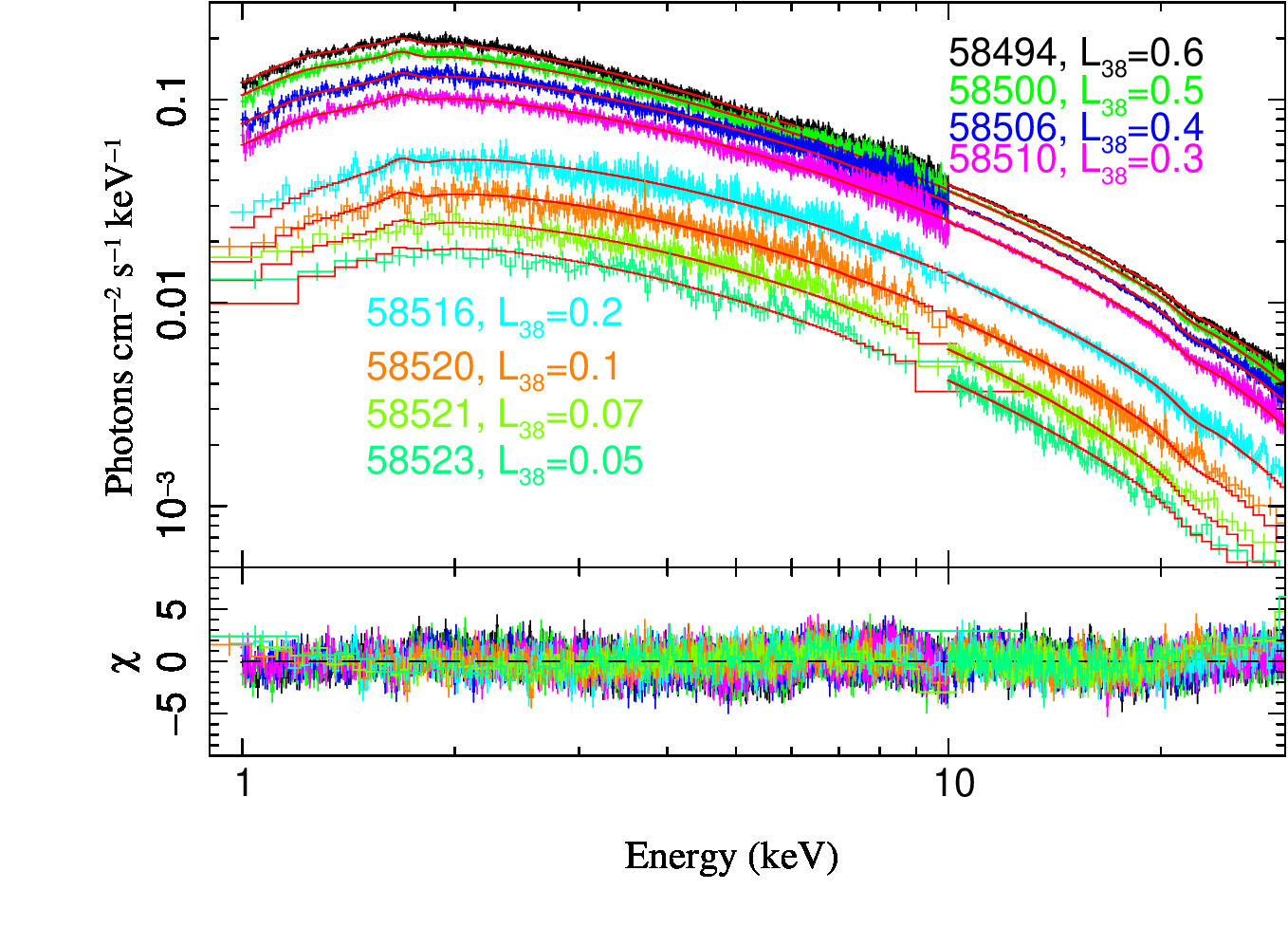}
	\caption{Example of the spectral evolution of Swift J0243 during the
	spin-up/spin-down transition at different times/luminosities.
} 
\end{figure}

As can be seen from Table 1, 
for the spectrum at $L_{38}=0.6$ to that at $L_{38}=0.07$, the fitted absorption 
column changes from $0.6\times10^{22}$\,cm$^{-2}$ 
to $0.72\times10^{22}$\,cm$^{-2}$, the fitted photon 
index changes from 0.8 to 0.5, and the cutoff energy changes from 15.5 keV to 
9.4 keV. The changing trend of the photon index and the cutoff energy is continuous, 
except for those value at the lowest luminosity of $L_{38}=0.05$.
The lower the luminosity, the harder the spectra, and the smaller the cutoff
energy.

\section{Discussion and conclusion}

We performed a detailed study of the spin-down accretion regime of Swift J0243 
after its giant outburst. The great brightness of Swift J0243 
makes its quiescence period observable and thus allows measurement
of its spin frequency during the quiescent spin-down regime.

Its spin frequencies after the giant outburst (after MJD 58170) 
show a linear decreasing trend, superposed with some variations due to 
the minor outbursts. 
The linear spin-down rate during the period of lowest luminosity (MJD 58190-58290)
is about $-1.9\times10^{-12}$ Hz/s. 
This spin-down rate is likely dominated by the braking torque 
due to the field threading of the accretion disk outside the co-rotation radius.
As the dipole field decreases as $r^{-3}$, the braking 
torque is mainly dominated by the threading near the co-rotation radius ($R_c$). 
Taking the braking torque as $\tau_{b}=-\mu^2/9R_c^3$ \citep{Wang95,Rap04}
and an accretion torque of $\dot{m}\sqrt{GMR_c}$ 
(assuming there exist accretion disk terminating at $R_c$), 
one can infer a dipole magnetic field of $B\sim1.75\times10^{13}$ G 
from $2\pi I\dot{\nu}=\dot{m}\sqrt{GMR_c}+\tau_{b}$, assuming a neutron star
of 1.4 $M_\odot$ with a radius of 10 km. 
Such a dipole field is similar to the value inferred from the spin-up
torque at higher luminosity with a radiation-pressure-dominated (RPD) disk
model \citep{Liu22}, and  
is consistent with the cyclotron line of Swift J0243 reported recently 
by \citet{Kong22}. It indicates that the dipole field of
Swift J0243 is similar to the magnetic field measured near the neutron star.


The observed $\dot{\nu}-L$ relation around the transition regime of 
spin-up and spin-down shows an offset between the later period of the giant 
outburst (around MJD 58160) and the last outburst around MJD 58500.
The zero point of $\dot{\nu}=0$ is around
$\sim0.15-0.2\times10^{38}$erg\,s$^{-1}$ for the giant outburst,
while it is around $\sim0.33\times10^{38}$erg\,s$^{-1}$ for the last outburst.
Such a phenomena is in contrast to the expectation of a simple torque 
model, which generally predicts a single $\dot{\nu}-L$ relation 
and a single point of zero torque. One possible reason is that 
for different outburst, the angular momentum of accreting
material may have different normal direction, which may lead to 
a different accretion torque for similar accretion rate.

\begin{figure}
	\hspace{-0.5cm}
	\includegraphics[width=3.5in]{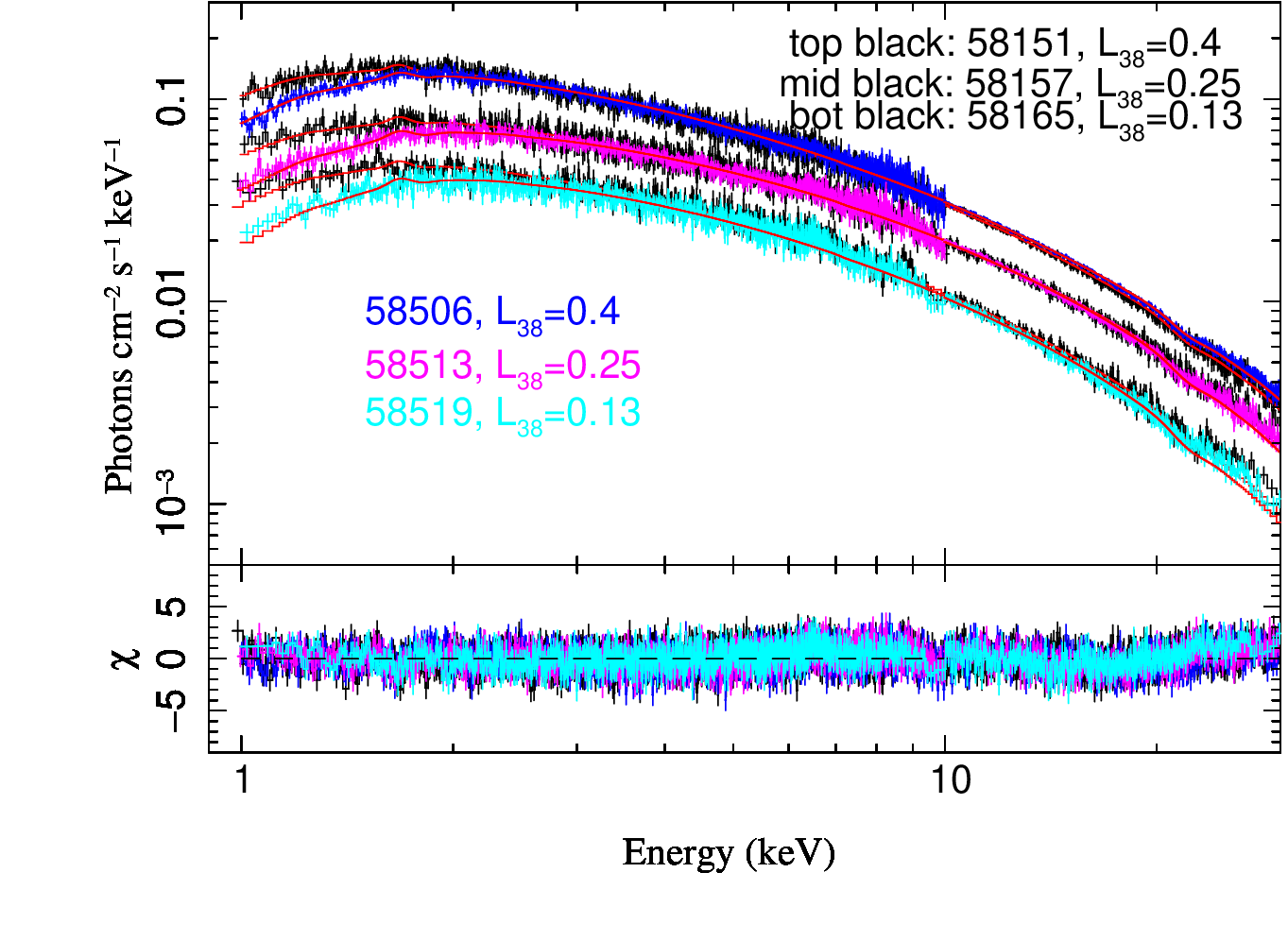}
	\caption{Comparison of the spectra extracted at similar luminosities 
	from the giant outburst (MJD 58151, 58157, and 58165, black) 
	and the last outburst (MJD 506, 58513, and 58519, colored).
} 
\end{figure}

\begin{figure}
	\hspace{-0.5cm}
	\includegraphics[width=3.5in]{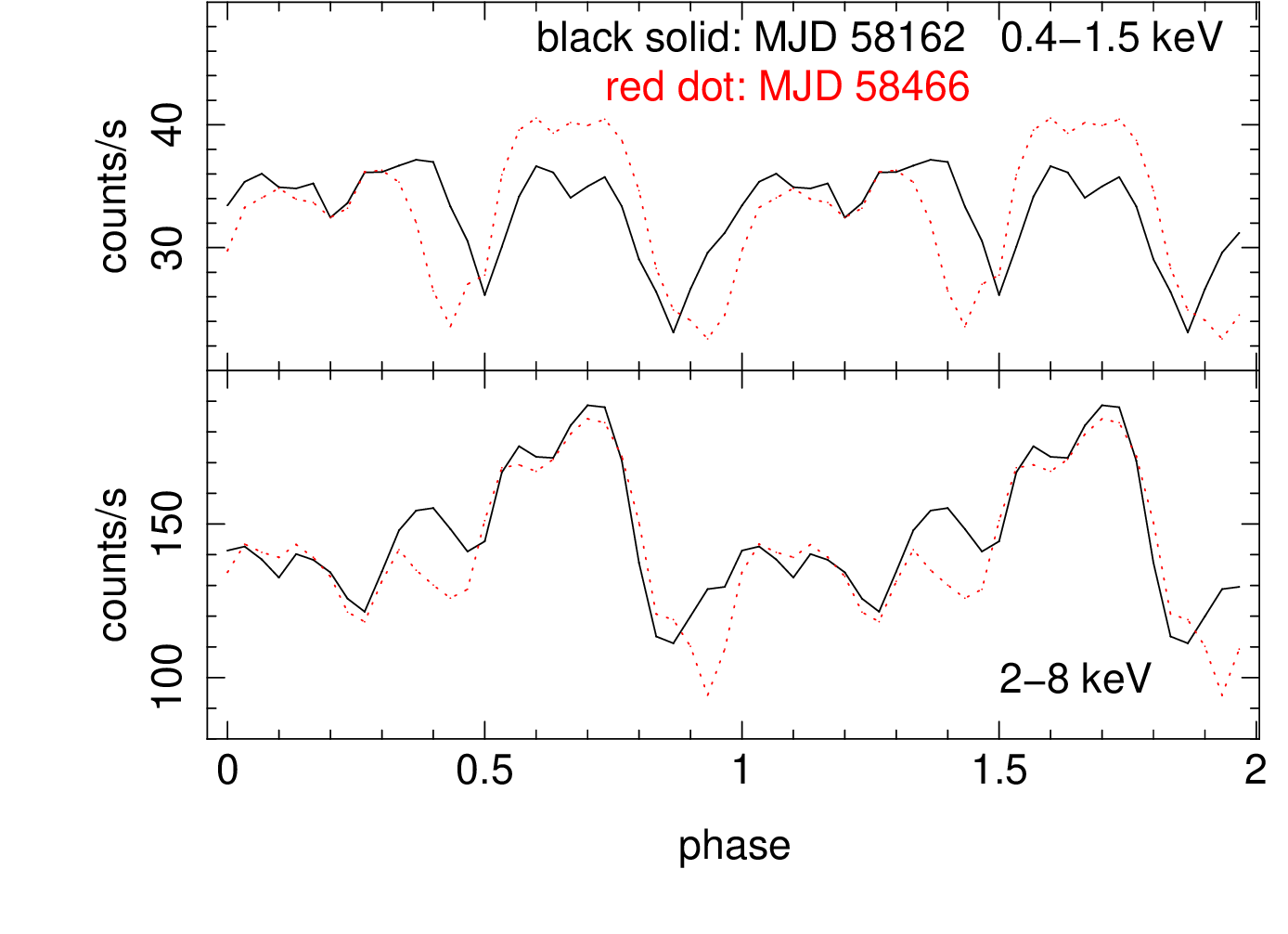}
	\caption{Comparison of the pulse profiles extracted from \nic data 
at a similar luminosity
of $L_{38}\sim0.22$ from the giant outburst (black solid, MJD 58162) and 
	the last outburst (red dot, MJD 58466) in 0.4-1.5 keV (top) and 2-8 keV (bottom). 
} 
\end{figure}

As pointed out by our anonymous referee, the two different zero-torques 
happen in different types of outbursts (the one around MJD 58160 is giant 
outburst, while the other one is from minor outbursts), which may have different 
accreting flows, and the difference of flows may be reflected in the spectrum and/or
pulse profile. In Figure 6, we present the spectra from both the giant outburst 
and the last outburst at similar luminosities of $L_{38}\sim0.13$, 0.25, and 0.4,
with the fitting results listed in the bottom part of table 1.
As can be seen, all three spectra of the giant outburst below 2 keV 
are systematically higher than those of the last outburst. 
The fitted absorption column densities of the giant outburst are 
typically less than those of the last outburst by $\sim0.2\times10^{22}$\,cm$^{-2}$.
It indicates there is less material between the emitting region of
the neutron star and us for the giant outburst than for the last outburst.

The pulse profiles in 0.4-1.5 keV and 2-8 keV for two observations at a similar 
luminosity ($L_{38}\sim0.22$) from the giant outburst (MJD 58162) and 
the last outburst (MJD 58466) are presented in Figure 7.
These profiles are in the middle state between the spin-up and the low spin-down
regime and are composed of three peaks. 
While the 2-8 keV profiles look quite similar, the 0.4-1.5 keV ones 
do show significant differences. The 0.4-1.5 keV main peak (around phase 0.7) 
of MJD 58162 is 
less significant than that of MJD 58466, while the two peaks around phase 
0.2 of MJD 58162 are broader than that of MJD 58466. 
The differences of the spectra and low-energy pulse profiles at a similar 
luminosity between 
the giant outburst and the last outburst indicate that the accreting 
flows are likely different for different types of outburst, and 
the different flows can lead to different points of zero-torque.

The $\dot{\nu}-L$ relation during the spin-down regime shows a complex 
behavior. During the peak period of the minor outburst, the luminosity 
is relatively high and the $\dot{\nu}$ is relatively large, close to zero, 
as expected. On the other hand, during some lowest flux periods, 
the $\dot{\nu}$ is also close to zero, which is unexpected.
This anomalous $\dot{\nu}$ behavior in the lowest spin-down regime may imply 
that there is much less material in this regime, or some other unknown reason.
Further study is needed to reveal the real physical origin. 

The pulse profile of Swift J0243 also shows significant evolution during 
the spin-up/spin-down transition. The profile is double-peaked 
in the spin-up state, but it changes to a single broad peak
around the luminosity of $L_{38}\sim0.04-0.08$ during the spin-down regime. 
In-between, the profile shows a transitional shape. 
In order to provide a much broader peak 
in the low spin-down regime, the 
emission beam/region of the neutron star should be larger for the low spin-down 
regime than that for the spin-up regime. This may reflect the more 
stochastic nature of the accreting flow during the spin-down regime.

The spectral characteristic of Swift J0243 during the spin-down regime 
is continuous and gradual with decreasing luminosity, except at the lowest
luminosity around $L_{38}\sim0.05$. The changing trend is similar to those below
an Eddington luminosity \citep[][]{Kong20}:
both the photon index and cutoff energy decreases with decreasing luminosity.
Such a behavior is similar to that of KS 1947+300 and EXO 2030+375, as reported 
by \citet{Rei13}. It is interesting to note that the decreasing trend of 
photon index and cutoff energy of KS 1947+300 and EXO 2030+375 reversed 
below $0.2$ and $1\times10^{37}$erg\,s$^{-1}$, respectively, and 
Swift J0243 shows a reversing trend around $0.5\times10^{37}$erg\,s$^{-1}$.

We note that the pulsation of Swift J0243 is still detected 
by \nic observations even when the luminosity 
is much lower ($\sim10^{35}$erg\,s$^{-1}$), 
as first illustrated by a \nus observation on MJD 58557 \citep{Dor20}.
The pulse profiles during these extremely low states
are also a single broad peak. Nevertheless, the signal is 
too low to provide a reasonable estimation of $\dot{\nu}$.
Based on the observed pulsation around $10^{34}$erg\,s$^{-1}$, some 
previous work \citep[e.g.][]{Dor20,Kong22}
suggested that Swift J0243 has not reached a propeller state at such a 
low luminosity and inferred a weak dipole field.
Nevertheless, the expectation of complete shutoff of accretion and pulsed emission 
during the propeller regime could be too ideal \citep[e.g.][]{ST93, Rap04}.
For example, many accreting millisecond X-ray pulsars show pulsed emission 
when the accretion rate is expected in the propeller regime
\citep[e.g.][]{Arc15,Pap15,San20}.
MHD simulations showed that accretion is still possible even when the 
magnetosphere radius is 5 times the co-rotation radius \citep[e.g.][]{Rom18}.
They proposed two reasons why accretion is possible when the magnetosphere
rotates more rapidly than the inner disk: only the closed part of 
the magnetosphere rotates more rapidly than the inner disk and the 
process is non-stationary.

Our results show that the accretion process is still ongoing during the 
low spin-down regime of Swift J0243, when the accretion torque is much lower than the 
braking torque. Many torque models predict a zero torque when 
the magnetosphere radius is close to the co-rotation radius 
\citep[e.g.][]{Wang95,KR07}. 
If this prediction is kind of true, 
the zero-torque luminosity ($L_{38}\sim0.33$) implies that when the luminosity
is 10 times smaller ($L_{38}\sim0.03$), 
the magnetosphere radius would be about 2 times the co-rotation radius. 
That is, accretion is still ongoing when the magnetosphere radius 
is a little larger than the co-rotation radius (the propeller regime).
The mechanism of accreting millisecond X-ray pulsar mentioned above may also 
apply to Swift J0243.
The accretion geometry during this 
low spin-down regime is most likely changed, as 
evidenced by the change of the pulse profile. 
Future more sensitive X-ray observatories will provide much great detail 
of the spin-down accretion regime of X-ray pulsars.

\section*{Acknowledgements}
We thank our referee for helpful comments and suggestion of the effects of different 
accretion flows and Zhang Shuang-Nan for helpful discussions. 
This research is supported by National Natural Science Foundation of China (U1938113)
and the Scholar Program of Beijing Academy of Science and Technology (DZ BS202002).
This research useed data obtained with Nicer, Fermi/GBM, Swift, and HXMT.


%
\bibliographystyle{mn2e}

\end{document}